\documentclass[twocolumn, times, tighten,twocolappendix]{aastex631}
\usepackage{graphicx,multirow,hyperref,url,color,xspace}
\usepackage{amsmath,amssymb}

\newcommand{\msun}{{\rm M}_{\sun}}

\newcommand{\integral}{{\textit{INTEGRAL}}\xspace}

\newcommand{\g}{$\gamma$}

\newbox\grsign \setbox\grsign=\hbox{$>$} \newdimen\grdimen \grdimen=\ht\grsign
\newbox\simpropbox
\setbox\simpropbox=\hbox{\raise.5ex\hbox{$\propto$}\llap
 {\lower.5ex\hbox{$\sim$}}}\ht2=\grdimen\dp2=0pt

\defcitealias{Zdziarski22}{Z22}

\begin{document}

\title{What are the Composition and Power of the Jet in Cyg X-1?}
\shorttitle{The Jet of Cyg X-1}
\author[0000-0002-0333-2452]{Andrzej A. Zdziarski}
\affiliation{Nicolaus Copernicus Astronomical Center, Polish Academy of Sciences, Bartycka 18, PL-00-716 Warszawa, Poland; \href{mailto:aaz@camk.edu.pl}{aaz@camk.edu.pl}}
\author[0000-0002-1532-4142]{Elise Egron}
\affiliation{INAF-Osservatorio Astronomico di Cagliari, Via della Scienza 5, I-09047 Selargius, Italy}

\shortauthors{Zdziarski \& Egron}

\begin{abstract}
We calculate the electron-positron pair production rate at the base of the jet of Cyg X-1 by collisions of photons from its hot accretion flow using the measurement of its average soft gamma-ray spectra by the Compton Gamma Ray Observatory and INTEGRAL satellites. We have found that this rate approximately equals the flow rate of the leptons emitting the observed synchrotron radio-to-IR spectrum of the jet, calculated using an extended jet model following that of Blandford \& K{\"o}nigl. This coincidence shows the jet composition is likely to be pair-dominated. The same coincidences were found before in the microquasar MAXI J1820+070 and in the radio galaxy 3C 120, which shows that the considered mechanism can be universal for at least some classes of relativistic jets. Furthermore, we recalculate the jet power of Cyg X-1. The presence of pairs can strongly reduces the power in the bulk motion of ions, which then limits the parameter space at which the jet can power the $\sim$5-pc nebular structure present in its vicinity.
\end{abstract}

\section{Introduction}
\label{intro}

We study here the issue of the abundance of electron-positron (e$^\pm$) pairs in the jet of the microquasar Cyg X-1. While there are strong hints that extragalactic jets contain substantial numbers of e$^\pm$ pairs (e.g., \citealt{Ghisellini12, Pjanka17, Snios18, Sikora20, Liodakis22}), this is much less clear for jets in microquasars. In either case, the mechanisms producing the putative pairs remain uncertain.

A viable mechanism is e$^\pm$ pair production within the jet base by collisions of photons emitted by the accretion flow (e.g., \citealt{Henri91, B99_pairs, Levinson11, Aharonian17, Sikora20}). Figures 1 and 3 of \citet{Zdziarski22c} and \citet{Zdziarski22a}, respectively, give illustrations of the assumed geometry, and so we do not show it again here. The rate of the pairs produced at the base can be compared with the flow rate of non-thermal relativistic electrons emitting synchrotron emission far downstream in the jet. Such comparisons were done for the microquasar MAXI J1820+070 \citep{Zdziarski22a} and for the radio galaxy 3C 120 \citep{Zdziarski22c}, in which cases these two rates were found to be similar, and compatible with being equal.

\begin{figure*}
\centerline{\includegraphics[height=\columnwidth]{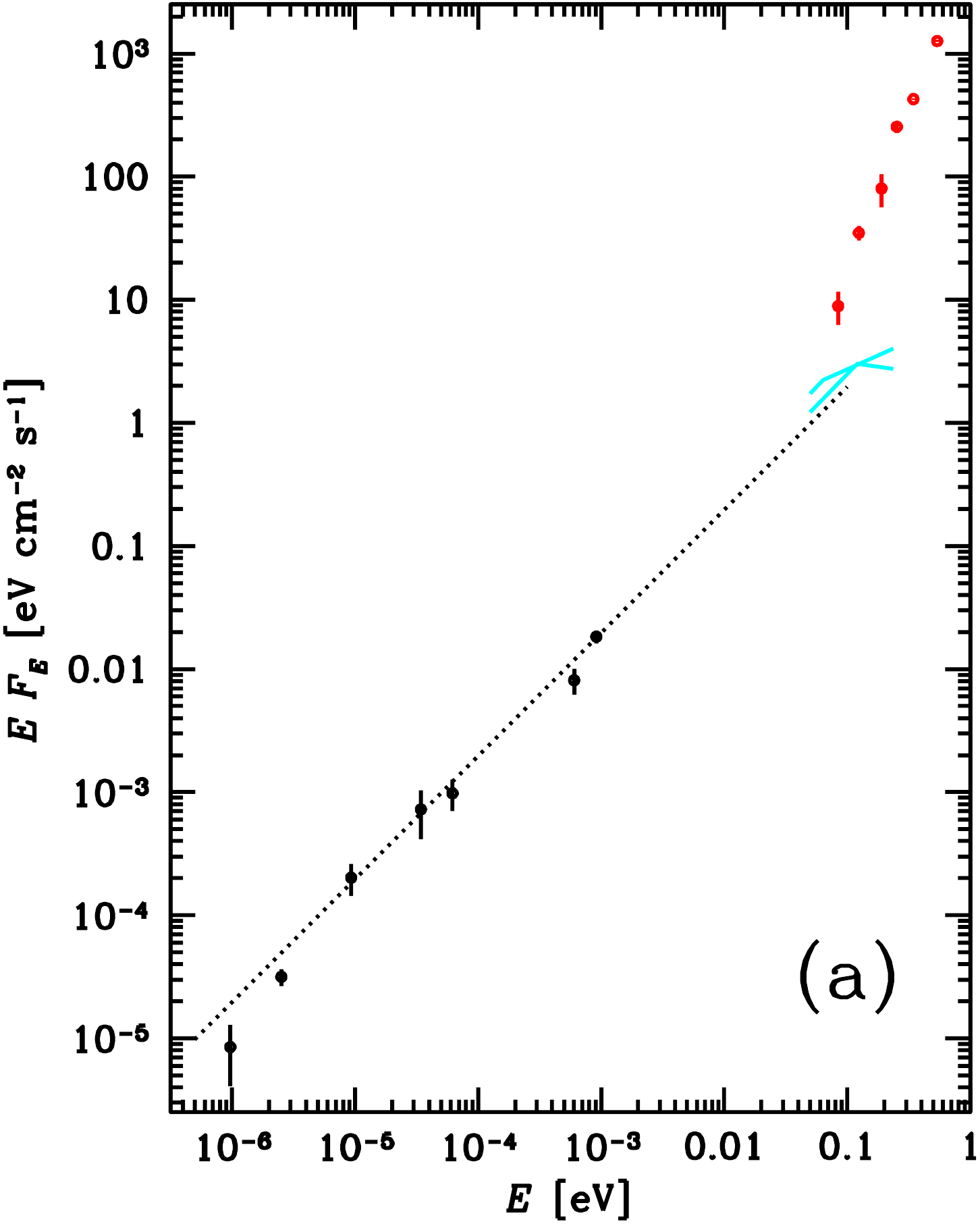}\hskip 1cm \includegraphics[height=\columnwidth]{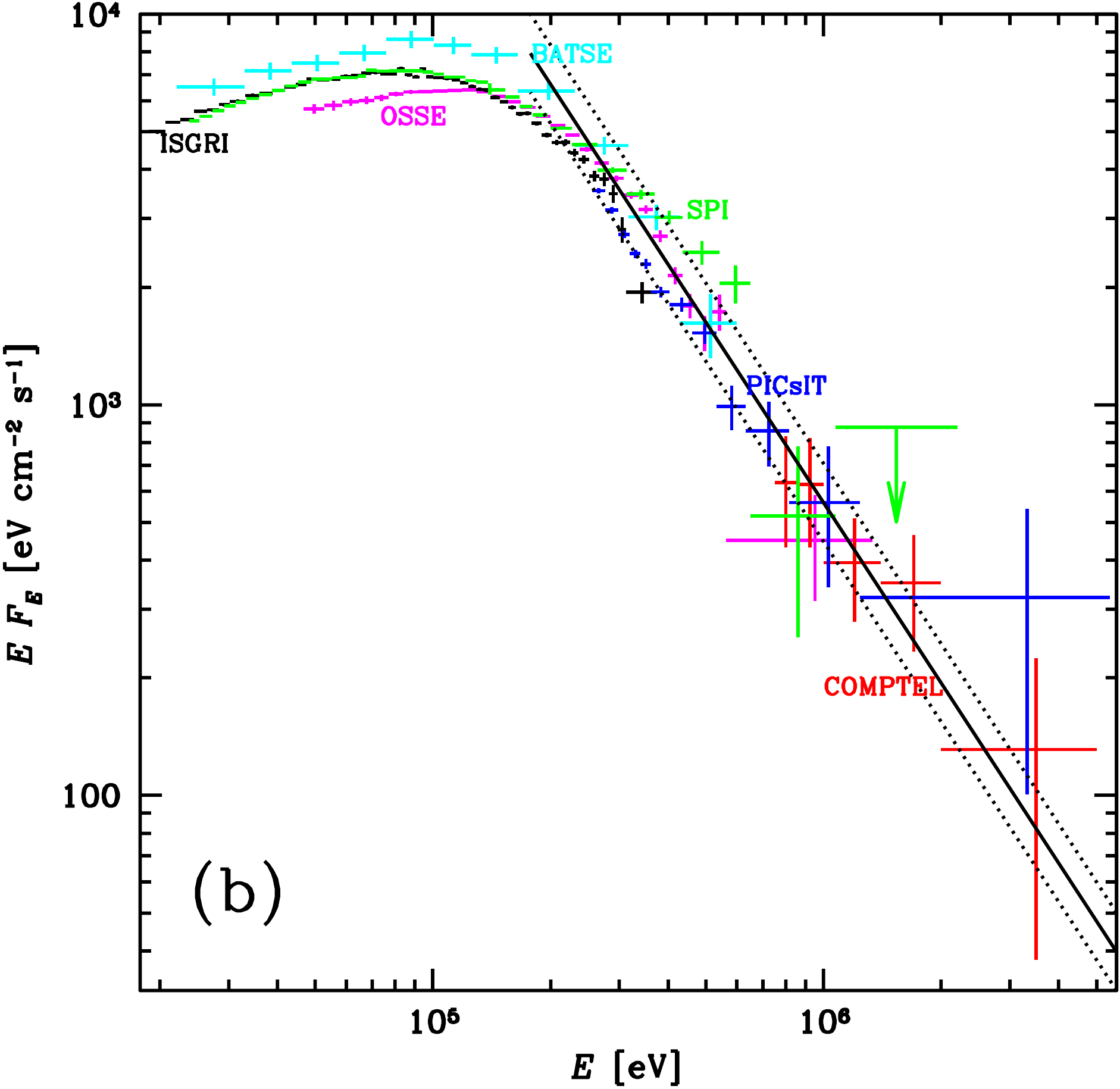}}
\caption{The spectra of Cyg X-1 in the hard state. (a) The radio to IR region. The black error bars give the fluxes from \citet{Pandey07} and \citet{Fender00}, and the red error bars are the IR fluxes \citep{Persi80,Mirabel96}, which are strongly dominated by the emission of the donor supergiant. The black dotted line gives a power-law fit to the 2.25--221\,GHz measurements from \citet{Fender00}, with $\alpha=0$ and $F_\nu=13$\,mJy. The 235 and 610\,MHz fluxes lie below that fit, which appears to be due to electron energy losses in the jet termination region \citep{Zdziarski14c}. The cyan symbols in the IR region show the broken-power law fit of \citet{Rahoui11}, claiming the possible position of the break frequency. (b) The hard X-ray to soft \g-ray average spectra measured by different instruments. The spectra from the OSSE, BATSE and COMPTEL detectors onboard Compton Gamma Ray Observatory are from \citet{McConnell02}, the spectra from the INTEGRAL ISGRI and PICsIT detectors are from \citet{ZLS12}, and that from the SPI detector is from \citet{JRM12}. The black solid line give a power law approximation to its high-energy tail with the spectral index of $\alpha_{\rm X}\approx -2.53$ and $E F_E(511\,{\rm keV})\approx 1.57\times 10^3$\,eV\,cm$^{-2}$\,s$^{-1}$. The dashed lines show an approximate uncertainty of the power-law normalization (with $\Delta\log_{10} F_E=\pm 0.1$).
}
 \label{spectra}
 \end{figure*}

Here, we perform such a comparison for Cyg X-1, which is probably the most studied microquasar, discovered in X-rays already in 1964 \citep{Bowyer65}. Its radio-to-mm spectrum in the hard state is flat \citep{Fender00}, with $\alpha\approx 0$ (defined by the energy flux of $F_\nu \propto \nu^\alpha$), see Figure \ref{spectra}(a). This is usually explained by the jet emission being partially synchrotron self-absorbed with both the distribution of non-thermal electrons and the magnetic energy flux maintained along the jet \citep{BK79}. This allows us to determine the rate of the total electron and positron flow through the jet based only on $F_\nu$ (independent of $\nu$ for $\alpha=0$) and the main jet parameters, see Appendix \ref{formulae}, without the need to specify either the location of the emission along the jet or the break frequency (above which the entire jet emission is optically thin). 

On the other hand, the rate of pair production within the jet base can be estimated based on the hard X-ray/soft \g-ray spectrum of the system. It has been well measured for Cyg X-1, see Figure \ref{spectra}(b). The spectrum above $\sim$200\,keV is relatively well approximated by a power law, which allows us to calculate this rate based on a fit by \citet{Svensson87}, see Appendix \ref{formulae}.

\section{Comparison of the flow rates}
\label{model}

The most recent determination of the binary parameters of Cyg X-1 is by \citet{Miller-Jones21}, who obtained the black-hole (BH) mass of $M_{\rm BH}\approx 21\pm 2\msun$, the mass of its supergiant donor of $M_1\approx 41\pm 7\msun$ (in agreement with \citealt{Ziolkowski14}) and the distance of $D\approx 2.2\pm 0.2$\,kpc. The binary inclination was given by them with very small uncertainties as $i_{\rm b}\approx 27.5\degr_{-0.6}^{+0.8}$. This, however, is in strong conflict with the inclination inferred from the X-ray polarization of Cyg X-1 measured by the Imaging X-ray Polarimetry Explorer \citep{Krawczynski22}, which yields much larger values of $\gtrsim\! 45\degr$ (see also \citealt{Ursini22}). It is likely that the viewing angle measured by the X-ray polarimetry corresponds to an inner hot accretion flow being inclined with respect to the binary axis but aligned with the BH spin axis. Furthermore, the polarization angle is aligned with the position angle of the jet, which suggest the jet is perpendicular to the inner hot flow, in which case its inclination is also $i\gtrsim\! 45\degr$. Thus, we consider in our analysis two different values for $i$.

The jet bulk Lorentz factor, $\Gamma$, and the opening angle, $\Theta$, were measured by \citet{Tetarenko19} as $\Gamma\approx 2.6^{+0.8}_{-0.6}$ and $\Theta\approx 0.4\degr$--$1.8\degr$. These values were obtained assuming $i\approx 27\degr$, while a lower $\Gamma$ and a larger $\Theta$ would be obtained for a larger value of $i$. Still, we adopt $\Gamma=2.6$ and $\Theta=1\degr$ as our default parameters. Based on the spectrum of \citet{Fender00}, shown in Figure \ref{spectra}(a), we use $F_\nu=13$\,mJy. The steady-state power-law index of accelerated electrons, $p$, in the jet of Cyg X-1 is unknown, but the dependence of the electron flow rate, $\dot N_{\rm e}$, on $p$ is relatively weak, see Equation (\ref{elflux}). We take $p=3$, approximately consistent with the models of \citet{Malyshev13} and \citet{Zdziarski14c}. As the default parameters, we take the energy density of the relativistic electrons equal to that of the magnetic field, $\beta_{\rm eq}=1$ (which approximately minimizes the jet power, see Appendix \ref{formulae}), the kinetic energy of ions as equal to that of the electrons, $k_{\rm i}=1$, and the number of quasi-thermal electrons below the minimum power-law Lorentz factor, $\gamma_{\rm min}$, equal to that in the power-law electrons, $k_{\rm e}=1$. The dependence on the maximum Lorentz factor, $\gamma_{\rm max}$, is negligible for $\gamma_{\rm max}\gg \gamma_{\rm min}$ and $p>2$; we assume $\gamma_{\rm max}=10^4$. At the above values and at $D=2.2$\,kpc, $i=27\degr$, we find the rate of the flow of leptons through the jet from Equation (\ref{elflux}),
\begin{equation}
\dot N_{\rm e}\approx 6.4\times 10^{39} \beta_{\rm eq}^{0.47}\!\left(\frac{\Gamma}{2.6}\right)^{2.7}\!\!\left(\frac{\Theta}{1\degr} \right)^{0.74}\!\!
\left(\frac{\gamma_{\rm min}}{3}\right)^{-1.5}\!\!{\rm s}^{-1}.
\label{Ndot1}
\end{equation}
This rate is also positively correlated with $i$, $\dot N_{\rm e}\propto (1-\beta\cos i)^\frac{14+6 p}{13+2 p} (\sin i)^\frac{2-2 p}{13+2 p}$, where the first term dominates the dependence. For $i=45\degr$, the numerical coefficient above increases to $1.8\times 10^{40}$.

We then compare those values with the pair production rate. The average hard X-ray/soft \g-ray spectra of the hard state of Cyg X-1 measured by various detectors are plotted in Figure \ref{spectra}(b). We have approximated the part of the spectra at $\gtrsim$200\,keV as a power law with the energy index of $\alpha_{\rm X}\approx -2.53$ (corresponding to the photon index of 3.53) and the normalization at 511\,keV of $E F_E\approx 1.57$\,keV\,cm$^{-2}$\,s$^{-1}$ with the uncertainty of $\Delta \log_{10} E F_E=\pm 0.1$. We then use the same formula as in \citet{Zdziarski22c}, given here by Equation (\ref{pprate}). This yields the rate of production of the sum of e$^+$ and e$^-$ of
\begin{equation}
2\dot N_+\approx 3^{+2}_{-1}\times 10^{40}\left(\frac{R_{\rm hot}}{10 R_{\rm g}}\right)^{-3}\left(\frac{R_{\rm jet}}{10 R_{\rm g}}\right)^2 {\rm s}^{-1},
\label{Nplus}
\end{equation}
where $R_{\rm g}=GM_{\rm BH}/c^2\approx 3\times 10^6$\,cm. The estimate of the radius of the base of the jet of $R_{\rm jet}\approx 10 R_{\rm g}$ follows from GRMHD simulations, see Fig.\ 3 in \citet{Zdziarski22a}, which used a simulation by \citet{Tchekhovskoy15} performed for the dimensionless spin parameter of $a_*=0.99$. On the other hand, the radius of the part of the hot disk with most of the high-energy emission is relatively uncertain. The used estimate corresponds to the case of a rapidly spinning BH, with $a_*\approx 1$, where we used twice the half-power radius, which can be approximated as $\approx [5 +28(1-a_*)]R_{\rm g}$ \citep{Fabian14}. If the spin is lower, both $R_{\rm hot}$ and $R_{\rm jet}$ will be larger. We see that, for likely jet parameters, it is larger than the obtained values for the lepton flow through the jet, see Equation (\ref{Ndot1}). We have also calculated the steady state optical depth of the pairs produced within the jet base, Equation (\ref{tau}), and found it to be $\tau_{\rm T}\lesssim 0.14$, where the upper limits corresponds to static pairs. Then, even a very modest advection velocity upstream the jet, with $\beta_{\rm adv}\gtrsim 0.03 c$, would result in the annihilation of the produced pairs being negligible. Thus, pair production by the accretion photons within the volume of the jet base appears to be entirely capable of providing enough leptons for the radio-mm synchrotron emission of the jet far away from the BH.

\section{The jet power}
\label{power}

The jet power in the kinetic energy of particles and the magnetic field at the parameters as used in Equation (\ref{Ndot1}) is, see Equation (\ref{pbe}),
\begin{equation}
P_{B{\rm e}}\approx 2.3\times 10^{35}\!\left(\frac{\Gamma}{2.6}\right)^{3.7}\!\!\left(\frac{\Theta}{1\degr} \right)^{0.74}\!\!
\left(\frac{\gamma_{\rm min}}{3}\right)^{-0.46}\!\frac{\rm erg}{\rm s}.
\label{pbe1}
\end{equation}
At $i=45\degr$, the numerical coefficient above increases to $6.4\times 10^{35}$. For $\beta_{\rm eq}\ll 1$ and $\gg 1$, $P_{B{\rm e}}$ is larger by the approximate factors $\beta_{\rm eq}^{-0.53}$ and $\beta_{\rm eq}^{0.47}$, respectively, which exponents follow from Equation (\ref{pbe}) at the assumed $p=3$.

The power in the bulk motion of cold ions at the parameters as used in Equation (\ref{Ndot1}) and for the hydrogen abundance of $X=0.5$ \citep{Miller-Jones21} is, see Equation (\ref{P_ion}),
\begin{align}
&P_{\rm i}\approx 2.0\times 10^{37} \beta_{\rm eq}^{0.47}\! \left(\frac{\Gamma}{2.6}\right)^{2.7}\frac{\Gamma-1}{1.6}\!\!\left(\frac{\Theta}{1\degr} \right)^{0.74}\times\\
&\left(\frac{\gamma_{\rm min}}{3}\right)^{-1.5}\left(1-\frac{2\dot N_+}{\dot N_{\rm e}}\right)
\frac{\rm erg}{\rm s}.
\label{Pi1}
\end{align}
At $i=45\degr$, the numerical coefficient above increases to $5.7\times 10^{37}$. 

If the $\sim$5\,pc-diameter structure discovered close to Cyg X-1 \citep{Gallo05} is indeed powered by one of its jets (which appears not fully certain, \citealt{Sell15}), the required total jet power is $\approx\! (1$--$3)\times 10^{37}$\,erg\,s$^{-1}$ \citep{Russell07}. This can put significant constraints on the jet parameter space, see also \citet{Malzac09}. 

We consider separately the cases of $\beta_{\rm eq}\ll 1$, $\approx$1 and $\gg$1. For $\beta_{\rm eq}\ll 1$ (a strongly magnetized plasma), $\dot N_{\rm e}$ is reduced by $\approx \beta_{\rm eq}^{0.5}$, which makes it even easier for pairs produced at the jet base to account for the produced leptons. $P_{\rm i}$ is then lowered by the same factors, and, if pairs dominate the flow, is lowered even more. Then, $P_{B{\rm e}}$ is larger by $\approx \beta_{\rm eq}^{-0.5}$, but, given its normalization, an extremely low $\beta_{\rm eq}$ would be required for $P_{B{\rm e}}$ to power the nebula. Furthermore, the 235 and 610 MHz data (Figure \ref{spectra}a) indicate that the relativistic electrons lose energy in the jet termination region, which would make the powering of the nebula by $P_{B{\rm e}}$ even more difficult. At $\beta_{\rm eq}\approx 1$, the jet can also be composed mostly of pairs. Then $P_{\rm i}\gg P_{B{\rm e}}$ unless there are about 100 pairs per ion. Since $P_{\rm i}$ decreases with the increasing pair abundance, it can power the nebula for modest pair abundances only. At $\beta_{\rm eq}\gg 1$, $P_{\rm i}\gg P_{B{\rm e}}$ in general, and the parameter space at which the base pair production can account for the synchrotron-emitting leptons is reduced, but it is still possible for some combinations of the parameters. The nebula can then be powered by $P_{\rm i}$. 

\section{Discussion}
\label{summary}

We consider here some effects relevant to our results. Detailed applications of the model of \citet{BK79} using the break energy of 0.15\,eV obtained by \citet{Rahoui11}, shown in Figure \ref{spectra}(a), lead to the determination of the distance of the 15\,GHz emission from the BH center, $z\sim 3\times 10^6 R_{\rm g}$ (e.g., \citealt{ZLS12, Zdziarski14c}), or $z\sim 3 A$, where $A\approx 4\times 10^{12}$\,cm is the separation between the stellar components. Such a distance is also in a qualitative agreement with the presence of a strong orbital modulation of the radio emission \citep{Pooley99}, which appears to be due to an orbital-phase dependent free-free absorption by the stellar wind of the donor. The large depth of the modulation at 15\,GHz of $\approx 30\%$ requires then $z$ to be comparable to $A$ \citep{SZ07, Zdziarski12}.  

On the other hand, \citet{Stirling01} and \citet{Rushton10} found that only $\sim$1/3--1/2 of the 8.4\,GHz emission is within the unresolved beam of VLBA, with the size of $\approx 10^{14}/\sin i$\,cm, and the remainder forms an extended jet up to the distance $\sim$5 times larger. At 15\,GHz, the VLBA beam is about twice smaller than that at 8.4\,GHz, while the resolved fraction is similar \citep{Rushton09, Zdziarski12}. An application of the partially self-absorbed model of \citet{BK79} yields then the location of the bulk of the 8.4\,GHz emission at $\approx 10^{14}/\sin i$\,cm \citep{Heinz06}, i.e., the distance about 30 times larger than that discussed above. This is in agreement with the result of \citet{Tetarenko19}, who found that 11\,GHz emission lags\footnote{While the 11\,GHz/X-rays lag appears relatively certain, we note that the relationship between the location and the emission frequency of $z\propto \nu^{-0.4}$ derived in \citet{Tetarenko19} using lags between different ratio frequencies cannot be reconciled with the model of \citet{BK79} and \citet{Konigl81} given the well-established spectral index of $\alpha\approx 0$. The two quantities together imply the electron density {\it increasing\/} fast with the distance, e.g., $a=-11.5$, $b=8.5$ (defined in Appendix \ref{formulae}) for $p=3$, obtained by combining equations (3) and (15) in \citet{Zdziarski22a}. Thus, either the obtained radio--radio lags are spurious or the model of partially self-absorbed synchrotron emission does not apply to Cyg X-1.} behind X-rays by $33.5^{+1.9}_{-1.7}$\,min, which corresponds to its source at $z\approx 2.5\times 10^{14}$\,cm (assuming $i\approx 30\degr$ and the jet velocity of $\beta\approx 0.9$). However, any orbital modulation due to wind absorption at this distance would be very tiny. A resolution of this discrepancy appears to require the presence of two dissipation regions in the jet, one at distances of the order of the separation, and one further away, as proposed in \citet{Zdziarski12}. The former can occur due to the jet-stellar wind interaction \citep{Perucho12, Yoon15, Yoon16}. The two components have similar fluxes, as follows from the comparison of the resolved and unresolved VLBA components. The hypothesis of the two dissipation regions can be verified by future studies of radio/X-ray time lags. In the case of the study of \citet{Tetarenko19}, there was a $\approx$2000-s gap in the X-ray coverage starting just at the onset of the main flare seen in the 11 and 9-GHz light curves. Then, any time lag present in the data shorter than several hundred s would be missing in that analysis. 

This would affect our results by lowering the lepton flow rate, $\dot N_{\rm e}$, by a factor of $\sim$2. This is because the electrons produced in the inner dissipation region could, after losing their energy, be re-accelerated in the outer dissipation zone. Thus, it would {\it increase\/} the parameter space within which the pairs produced by photon-photon collisions at the jet base can account for the synchrotron emission far upstream in the jet. Still, given the uncertainties of this process we do not take this into account in our formalism.

Another issue concerns the measured very high polarization degree (reaching unity at high energies) found in the \integral SPI data by \citet{Jourdain12}. If this is due to the synchrotron emission of the jet, it would be emitted close to the onset of the electron acceleration, which is inferred to be at $z\gg R_{\rm g}$ in Cyg X-1 (e.g., \citealt{Zdziarski14c}). Then, the emitted soft \g-rays would not have the density sufficient for efficient pair production. However, we consider that origin uncertain, and in this work we assume the soft \g-rays are emitted within the accretion flow. This is agreement with the modelling by \citet{McConnell02}, and strongly suggested by the shape of the spectrum, with the tail smoothly joining the emission at lower energies.

\section{Conclusions}

The observed hard X-ray/soft \g-ray emission of Cyg X-1 in its hard state extends to several MeV, forming a steep high-energy tail originating close to the peak (in $E F_E$) of the X-ray emission at lower energies, see Figure \ref{spectra}(b). The origin of this X-ray/\g-ray spectrum is likely to be Comptonization of some soft seed photons by hybrid (thermal with a high-energy tail) electrons within the hot accretion flow \citep{McConnell02}. We have found that e$^\pm$ pair production by collisions of the photons from the photon tail can readily supply enough electrons and positrons to account for the radio-to-mm emission of the jet, under the assumption that the latter originates from the partially-self-absorbed synchrotron mechanism, which is the standard explanation for jet spectra with $\alpha\sim 0$ \citep{BK79}. Thus, the composition of the jet in Cyg X-1 can be dominated by e$^\pm$ pairs. Such conclusions were achieved before for the microquasar MAXI J1820+070 \citep{Zdziarski22a} and the radio galaxy 3C 120 \citep{Zdziarski22c}, which shows that the considered mechanism can be quite universal. 

The presence of pairs reduces the density of the ions in the jet, and, consequently, the power in the ion bulk motion (dominating the total power in the absence of pairs) can be strongly reduced. This makes it less likely that the jet of Cyg X-1 powers the nebula discovered by \citet{Gallo05}. However, no reliable value of the jet power vs.\ the pair abundance can be obtained due to the present uncertainty regarding the parameters of the jet (e.g., its equipartition parameter). 

\section*{Acknowledgements}

We thank the referee for valuable comments. We acknowledge support from the Polish National Science Center under the grant 2019/35/B/ST9/03944. This research benefitted from discussion at Team meetings at the International Space Science Institute (Bern).   

\appendix
\section{Formulae}
\label{formulae}

We give here the explicit formulae for the electron flow rate and the jet power based on the formulation of the \citet{BK79} and \citet{Konigl81} model presented in \citet{Zdziarski19b}, see also \citet{Zdziarski22a}. We consider power-law dependencies of the electron number density and the magnetic field strength, $n(z,\gamma)= n_0(z/z_0)^{-a} \gamma^{-p}$ (where $\gamma$ is the electron Lorentz factor), $B(z)=B_0 (z/z_0)^{-b}$, respectively. Here $z_0$ is the distance of the onset of the acceleration. We assume the canonical values of $a=2$, $b=1$, for which $\alpha=0$. In this case, the rate of the flow of leptons (electrons and positrons) and the jet power are constant through the jet, and they are thus independent of the value of $z_0$. We give a formulation also suitable for extragalactic sources, including the dependence on the cosmological redshift, $z_{\rm r}$, and with the distance to the source, $D$, being the luminosity distance.  The flow rate can be derived as
\begin{align}
&\dot N_{\rm e}= (1+k_{\rm e})
\frac{3^{\frac{18+2 p}{13+2 p}} c^{\frac{23+2 p}{13+2 p}} (m_{\rm e}/4)^{\frac{5}{13+2 p}}f_N \beta\Gamma }{e^2 \delta^{\frac{14+6 p}{13+2 p}}} \times
\nonumber\\
&\left[\frac{\beta_{\rm eq}}{(1+k_{\rm i})(f_E-f_N)}\right]^{\frac{3+2 p}{13+2 p}}
\left[\frac{\pi C_2(p)}{\sin i}\right]^{\frac{2 p-2}{13+2 p}}\times
\label{elflux}\\
&\left[\frac{5 F_\nu  D^2\tan\Theta} {C_1(p)(1+z_{\rm r})\Gamma_{\rm E}\left(\frac{p-1}{p+4}\right)}\right]^{\frac{8+2 p}{13+2 p}}, \nonumber
\end{align}
where $k_{\rm e}$ is the relative contribution of quasi-thermal electrons below $\gamma_{\rm min}$, $e$ is the electron charge, $C_{1,2}$ are functions of $p$ defined, e.g., in equations (8) and (9) of \citet{Zdziarski22a}, respectively, $\Gamma_{\rm E}$ is the Euler Gamma function, $\delta$ is the Doppler factor,
\begin{align}
&\delta=\frac{1}{\Gamma(1-\beta\cos i)},\quad \beta_{\rm eq}={n_0 m_{\rm e} c^2 (1+k_{\rm i})(f_E- f_N)\over B_0^2/8\pi},\label{betaeq}\\
&f_E\equiv \begin{cases} {\gamma_{\rm min}^{2-p}-\gamma_{\rm max}^{2-p}\over p-2}, &p\neq 2;\cr
\ln {\gamma_{\rm max}\over \gamma_{\rm min}},& p=2,\cr
\end{cases}\quad
f_N\equiv \frac{\gamma_{\rm min}^{1-p}-\gamma_{\rm max}^{1-p}}{p-1},
\label{fe_fn}
\end{align}
and $k_{\rm i}$ is the fractional contribution to the total kinetic energy density in particles other than the power-law electrons, i.e., ions and electrons below $\gamma_{\rm min}$. 

Then the jet power in leptons and magnetic field of both the jet and the counterjet is also constant along the distance from the BH,
\begin{align}
&P_{B{\rm e}}= 
\frac{2^{\frac{3+2 p}{13+2 p}} 3^{\frac{5}{13+2 p}} c^{\frac{49+6 p}{13+2 p}} m_{\rm e}^{\frac{18+2 p}{13+2 p}}\beta\Gamma^2 (3+2\beta_{\rm eq}) }{e^2 \delta^{\frac{14+6 p}{13+2 p}}} \times
\nonumber\\
&\left[\frac{(1+k_{\rm i})(f_E-f_N)}{\beta_{\rm eq}}\right]^{\frac{10}{13+2 p}}
\left[\frac{\pi C_2(p)}{\sin i}\right]^{\frac{2 p-2}{13+2 p}}\times
\label{pbe}\\
&\left[\frac{5 F_\nu  D^2\tan\Theta} {C_1(p) (1+z_{\rm r})\Gamma_{\rm E}\left(\frac{p-1}{p+4}\right)}\right]^{\frac{8+2 p}{13+2 p}}. \nonumber
\end{align}
This power is minimized for $\beta_{\rm eq}=15/(3+2 p)$. The usable power in the bulk motion of cold ions is
\begin{equation}
P_{\rm i}=\mu_{\rm e} m_{\rm p}c^2 (\Gamma-1)\left(\dot N_{\rm e}- 2\dot N_+\right),\label{P_ion}
\end{equation}
where $\mu_{\rm e}=2/(1+X)$ is the mean electron molecular weight, $X$ is the hydrogen mass content and $m_{\rm p}$ is the proton mass. The term $(\Gamma-1)$ takes into account the fact that rest-energy flow through the jets is provided by the accretion flow. Then, the minimum of the total jet power depends on the unknown pair abundance, and we thus do not give it here.

For completeness, we also provide the used formula for pair production rate by the accretion photons within the jet base, see equations (A10--A11) of \citet{Zdziarski22c} and \citet{Svensson87},
\begin{align}
&\dot N_+\approx 2\pi R_{\rm jet}^2 R_{\rm hot} n_1^2 \sigma_{\rm T} c\frac{\ln(E_{\rm c}/511\,{\rm keV})} {(1-\alpha_{\rm X})^{5/3} (2-\alpha_{\rm X})},
\label{pprate}\\
&n_1\approx F_E(511\,{\rm keV})\frac{4\pi D^2}{2\pi R_{\rm hot}^2 c},
\label{n1}
\end{align}
where $n_1$ is the differential photon density at 511\,keV (in the source frame), above the hot disk, $F_E$ and $E$ are assumed to have the same energy unit (e.g., eV), $R_{\rm hot}$ and $R_{\rm jet}$ are the characteristic radii of the hot accretion flow and the jet, respectively, $\sigma_{\rm T}$ is the Thomson cross section, and $E_{\rm c}$ is the upper cutoff of the photon power law, which we assume $\approx$3\,MeV. The upper limit on the optical depth of the pairs within the jet base can be roughly estimated as \citep{Zdziarski21c}
\begin{equation}
\tau_{\rm T,max}\approx \frac{14\ln(E_{\rm c}/511\,{\rm keV})\sigma_{\rm T} F_E(511\,{\rm keV}) D^2}{(1-\alpha_{\rm X})^{5/3} (2-\alpha_{\rm X})},
\label{tau}
\end{equation}
which is obtained by equating the local pair production and pair annihilation rates. The rate of pair advection will be faster than that of the annihilation for $\beta_{\rm adv}>(3/16)\tau_{\rm T}$ \citep{Zdziarski21c}. Thus, if $\tau_{\rm T}$ is low, even a very low $\beta_{\rm adv}$ will prevent substantial annihilation within the jet base. Then, most of the pairs will be advected upstream, where further annihilation is completely negligible. 

\bibliography{../../allbib}{}
\bibliographystyle{../../aasjournal}

\end{document}